\documentstyle[aps,multicol]{revtex}
\begin{document}

\title{Bell inequalities for arbitrarily high dimensional systems}
\author{Daniel Collins$^{1,2}$, Nicolas Gisin$^3$, Noah Linden$^4$, 
Serge Massar$^5$, Sandu Popescu$^{1,2}$ }
\address{$^1$ H. H. Wills Physics Laboratory, University of Bristol,
Tyndall Avenue, Bristol BS8 1TL, UK\\
$^2$ BRIMS, Hewlett-Packard Laboratories, Stoke Gifford, Bristol BS12
6QZ, UK
\\
$^3$ Group of Applied Physics, University of Geneva, 20, rue de
l'Ecole-de-M\'edecine, CH-1211 Geneva 4, Switzerland
\\
$^4$ Department of Mathematics, Bristol University, University Walk,
Bristol BS8 1TW, UK\\
$^5$ Service de Physique Th\'eorique,
Universit\'e Libre de Bruxelles, CP 225, Bvd. du Triomphe, B1050
Bruxelles, Belgium.\\
}
\date{18-07-01}

\maketitle

\begin{abstract}
We develop a novel approach to Bell inequalities based on a
constraint that the correlations exhibited by 
local variable theories must satisfy. This is used to construct a
family of Bell inequalities for bipartite quantum systems of
arbitrarily high
dimensionality which are strongly resistant to noise. In particular 
our work gives an analytic description of previous numerical results, and
generalizes them to arbitrarily high dimensionality.
\end{abstract}

\vspace{0.2cm}

\begin{multicols}{2}

One of the most remarkable aspect of quantum mechanics is its predicted
correlations.
Indeed, the correlations between outcomes of measurements performed on
systems composed
of several parts in an entangled state have no classical analog. The most
striking aspect of this
characteristic feature of quantum physics is revealed when the parts are
spatially separated:
no classical 
theory based on local variables can reproduce the quantum correlations.
Historically, this
became known as the EPR paradox and was formulated in terms of measurable
quantities by Bell \cite{Bell} and
by Clauser, Horn, Shimony and Holt \cite{CHSH} as the nowadays
famous inequalities. Other aspects
of quantum correlation were analyzed in the form of paradoxes, like, e.g.
Schr\"odinger's cat and
the measurement problem. In recent years, these paradoxical aspects
have been  
overthrown by a more
effective approach: {\it let's exploit ``quantum strangeness'' to perform
tasks that are classically impossible}
has become the new leitmotiv! From this ``conceptual revolution'', 
the field of
quantum information emerged.
Old words became fashionable, like ``entanglement''. Old questions
were revisited, like the classifications of
quantum correlation. 

The variety of known partial results, in particular
about entanglement measures, makes it
today obvious that there is no one-parameter classification of
entanglement. 
This letter
concerns classifications related to
what is called quantum non-locality, i.e. the impossibility to reproduce
quantum correlations with theories
based on local variables (often called local realistic theories).      
Specifically we develop a powerful new approach to Bell
inequalities which we then use to write several families of Bell
inequalities for higher dimensional systems.

Local variable theories cannot exhibit arbitrary correlations. 
Rather 
the conditions these correlations must obey can always be written as
inequalities (the Bell inequalities) 
which the joint probabilities of
outcomes must satisfy.
Our approach to Bell inequalities is based on a logical
constraint the correlations must satisfy in
the case of local variable theories.
In order to introduce this
constraint, 
let us suppose that one of the parties, Alice, can carry 
out two possible measurements, $A_1$ or $A_2$, and that the other
party, Bob, can carry out two possible measurements, $B_1$ or $B_2$.
Each measurement may have $d$ possible outcomes:
$A_1, A_2, B_1, B_2 =0,\ldots, d-1$.
Without loss of generality 
a local variable theory can be described by $d^4$ probabilities
$c_{jklm}$ (${j,k,l,m}= 0,\ldots, d-1$)
that Alice's local variable ($jk$) specifies that
measurement $A_1$ gives outcome $j$ and that
measurement $A_2$ gives outcome $k$; and that 
Bob's local variable ($lm$) specifies that
measurement $B_1$ gives outcome $l$ and that
measurement $B_2$ gives outcome $m$. (In this formulation
Alice and Bob's strategy is deterministic since it is completely
determined by the value of their variables $jk$ and $lm$. Any non
deterministic local theory can be rephrased in the above way by
incorporating the local randomness in the probabilities $c_{jklm}$,
see for instance \cite{Percival}).
Since they are probabilities the $c_{jklm}$ are positive ($
c_{jklm}\geq 0$) and sum to one ($\sum_{jklm} c_{jklm} =1$).
The joint probabilities
take the form
$P(A_1=j, B_1=l) = \sum_{km} c_{jklm}$, and similarly for
$P(A_1=j, B_2=m)$, 
$P(A_2=k, B_1=l)$
and 
$P(A_2=k, B_2=m)$.

Let us consider a
particular choice of local variables $jklm$ (this choice occurs with 
probability $c_{jklm}$).  Since $A_1=j$, $A_2=k$, $B_1=l$,
$B_2=m$ we have
\begin{eqnarray}
r' &\equiv& B_1- A_1 = l-j \ ,\nonumber\\
s'&\equiv& A_2- B_1  = k-l\ ,\nonumber\\
t' &\equiv& B_2- A_2 = m-k \ ,\nonumber\\
u'&\equiv& A_1- B_2  = j-m\ .
\label{defff}
\end{eqnarray}
We see that the difference, $r'$, between $A_1$ and $B_1$ can be
freely chosen by choosing $j$ and $l$. Similarly the difference, $s'$,
between $B_1$ and $A_2$ and the difference, $t'$, between $A_2$ and $B_2$ can 
be freely chosen. But then the difference $u'$ between $B_2$ and $A_1$ 
is constrained since we necessarily have
\begin{equation}
r' + s'+t'+u'=0 \ .
\label{constraint'}
\end{equation}
Thus in a local variable theory the relation between three pairs of
operators can be freely chosen, but then the last relation is
constrained. 

This constraint plays a central role in our Bell inequalities. Indeed
they are written in such a way that their maximum
value can be attained only if this constraint is frustrated.
The simplest such Bell expression is 
\begin{eqnarray}
I &\equiv& P(A_1=B_1) +  P(B_1= A_2+1)\nonumber\\
& & + P(A_2=B_2) +  P(B_2=A_1)
\label{I}
\end{eqnarray}
where we have introduced the probability $P(A_a=B_b+k) $
that the measurements $A_a$ and $B_b$ have outcomes that differ, modulo 
$d$, by $k$:
\begin{eqnarray}
P(A_a=B_b+k)  \equiv \sum_{j=0}^{d-1} P(A_a=j, B_b= j + k  \mbox{ mod d}) \ .
\nonumber\\
\label{modulo}\end{eqnarray}
Because the difference between $A_a$ and $B_b$ is evaluated modulo
$d$, all the outcomes of $A_a$ and $B_b$ are treated on an equal
footing. As we see in eq. (\ref{I}) this symmetrization is the key to
reducing Bell inequalities to the logical constraint that is imposed
by local variable theories.
Indeed because 
of the constraint eq. (\ref{constraint'}) any choice of local
variables $jklm$ 
can satisfy only three of the relations appearing in
eq. (\ref{I}), eg. $A_1=B_1$, $B_1= A_2+1$, etc\ldots .
Hence $I(\mbox{local realism}) \leq 3$. On the other
hand non-local correlations can attain $I=4$ since they can satisfy
all 4 relations.

In the case of two dimensional systems the inequality  $I(\mbox{local
  variable}) \leq 3$ 
is equivalent to the CHSH inequality\cite{CHSH}. But the power of our
reformulation is already apparent since this inequality generalizes the
CHSH inequality to arbitrarily large dimensions.  In fact the above
formulation of the constraint imposed by local realistic theories
allows one to write in a unified way all previously known Bell
inequalities\cite{inpreparation}. It can also serve to write
completely new Bell inequalities and this is the subject of the
present article. 
Specifically we have generalised in a non trivial way (see
eqs. (\ref{I3}) and (\ref{Id}) below)  
the Bell expression (\ref{I}) to $d$ dimensional systems (for any $d\geq 2$).

One of the interests of these new Bell expressions is that they are
highly resistant to noise.
Indeed Bell inequalities are sensitive to the 
presence of noise and above a certain amount of noise the Bell
inequalities will cease to be violated by a quantum system. However it
has been shown
by numerical optimization \cite{1} 
that using higher dimensional systems can increase
the resistance to noise. The measurements that are carried out on
the quantum system in order to obtain an increased violation have been 
described analytically in \cite{2}. And an analytical proof of the
greater robustness of quantum systems of dimension 3 was given in
\cite{3}. One of the interests of our new Bell inequalities is that when
we apply them to 
the quantum state and measurement described in
\cite{2} for those dimensions ($d\leq 16$) for which a numerical
optimisation was carried out in \cite{2}, 
we obtain the same resistance to noise as in \cite{2}.

The first generalisation of the Bell expression
eq. (\ref{I}) is
\begin{eqnarray}
I_3 = &+& \left [ P(A_1=B_1) +  P(B_1= A_2+1) \right.\nonumber\\
 & &\left.\  + P(A_2=B_2) +  P(B_2=A_1) \right]
\nonumber\\
&  -& \left [ P(A_1=B_1-1) + P(B_1= A_2)  \right.\nonumber\\
 & & \ \left.  + P(A_2=B_2-1) +  P(B_2=A_1-1)\right]
\ .
\label{I3}
\end{eqnarray}

The maximum value of $I_3$ for non-local theories is 4 since a non
local theory could satisfy all 4 relations that have a $+$ sign in
(\ref{I3}). On the other hand for a local variable theory $I_3 \leq
2$. This should be compared to the constraint 
$I (\mbox{local variable})\leq 3$ for the expression
(\ref{I}). The origin of this difference is the $-$ signs
in (\ref{I3}). Indeed we have seen when analyzing (\ref{I}) that only
three of the relations with a $+$ sign can be satisfied by local
realistic theories. But if 3 relations with $+$ are satisfied in
(\ref{I3}), 
then
necessarily one relation with $-$ is also satisfied giving a total of
$I_3 =2$. Alternatively one can satisfy 2 relations with $+$ and two
relations with weight zero (if the dimension is larger than 2), once
more giving a total of $I_3=2$. 

For $d=2$ the inequality $I_3(\mbox{local variable}) \leq 2$ is equivalent to 
the inequality $I(\mbox{local variable}) \leq 3$ and therefore to the
CHSH inequality. But for $d\geq 3$ the inequality based on $I_3$ is 
not equivalent to that based on $I$. For the quantum measurement
described below (when $d \geq 3$) the inequality based
on $I_3$ (and its generalisations $I_d$ given below) is more robust
than that based on $I$. 

The Bell expression $I_3$ can be further generalised when the
dimensionality is greater than 3 by adding extra terms. 
The extra terms  in $I_d$ do not change the maximum
value attainable by local variable theories 
($I_d^{max}(\mbox{local variable}) =2$), nor do they change the
maximum value attainable by completely non local theories
($I_d^{max}=4$).
However these extra terms allow a better exploitation of the
correlations exhibited by quantum systems.

These new Bell expressions have the form:
\begin{eqnarray}
I_d &=& \sum_{k=0}^{[d/2] -1} \left ( 1 - { 2 k \over d-1} \right) \nonumber\\
& \Bigl(& + \left[  P(A_1=B_1 + k) +  P(B_1= A_2+k +1) \right.\nonumber\\
& &\left . \ \ 
+ P(A_2=B_2 + k ) +
  P(B_2=A_1+ k)\right]\nonumber\\
& & -\left[  P(A_1=B_1 - k-1) +  P(B_1= A_2-k) \right.\nonumber\\
& &\left . \  \ 
+ P(A_2=B_2 - k-1 ) +
  P(B_2=A_1- k-1)\right ] \Bigr)
\ .
\label{Id}
\end{eqnarray}
As mentioned above the maximum value of $I_d$ is 4. This follows
immediatly from the fact that the maximum weight of the terms in 
(\ref{Id}) is $+1$. And the maximum value of $I_d$ for local variable
theories is 2. We now prove this last result.

The proof consists of enumerating all the possible relations between
$A_1, B_1, A_2, B_2$
allowed by the constraints (\ref{constraint'}). 
This is most easily done by first 
changing 
notation. We do not use the coefficients $r', s', 
t',  u'$ defined in (\ref{defff}), but  use new coefficients $r, s,
t, u$ defined by the relation
\begin{eqnarray}
&A_1 = B_1 + r\ , \ B_1 = A_2 + s +1\ ,& \nonumber\\ 
&A_2 = B_2 + t\ , \ B_2 = A_1 + u\ ,&
\end{eqnarray} which obey the constraint
\begin{equation}
r + s + t + u + 1 = 0 \mbox{ mod } d \ .
\label{sum}
\end{equation}
Furthermore we restrict (without loss of generality) $r, s,t, u$ 
to lie in the interval
\begin{equation}
-[d/2] \leq r , s , t , u \leq [(d-1)/2]
\label{interval}
\end{equation}
With this notation the value of the Bell inequality for a given choice 
of $r, s, t, u$ is
\begin{equation}
I_d(r,s,t,u) = f(r) + f(s) + f(t) + f(u)
\label{If}
\end{equation}
where $f$ is given by
\begin{equation}
f(x) = \left\{
\begin{array}{ccc}
- {2 x \over d-1} + 1 & , &  x\geq 0 \\
 - {2 x \over d-1} - {d + 1 \over d-1} & , &  x <0
\end{array}
\label{f}
\right.
\end{equation}
We now consider different cases according to
the signs of $r, s,t,u$.
\begin{enumerate}
\item $r,s,t,u$ are all positive. Then (\ref{sum}) and (\ref{interval})
  imply that $r + s+ t+u =d-1$. Inserting into (\ref{If}) and using 
  (\ref{f}) one finds $I_d =2$.
\item Three of the numbers $r,s,t,u$ are positive, one is strictly negative.
Then (\ref{sum}) and (\ref{interval}) imply that
either $r + s+ t+u =d-1$ or $r + s+ t+u =-1$. 
Inserting into (\ref{If}) and using 
  (\ref{f}) one finds either $I_d =-2/(d-1)$ or $I_d = 2$.
\item Two of the numbers $r,s,t,u$ are positive, two are strictly negative.
Then (\ref{sum}) and (\ref{interval}) imply that $r + s+ t+u =-1$. 
Inserting into (\ref{If}) and using 
  (\ref{f}) one finds $I_d =-2/(d-1)$.
\item  One of the numbers $r,s,t,u$ is positive, three are strictly negative.
Then (\ref{sum}) and (\ref{interval}) imply that either $r + s+ t+u =-1$ 
or $r + s+ t+u =-d-1$. 
Inserting into (\ref{If}) and using 
  (\ref{f}) one finds either $I_d =-2 (d+1) / (d-1)$ or $I_d =-2/(d-1)$.
\item  The numbers $r,s,t,u$ are all strictly negative.
Then (\ref{sum}) and (\ref{interval}) imply that $r + s+ t+u =-d-1$. 
Inserting into (\ref{If}) and using 
  (\ref{f}) one finds $I_d =-2 (d+1)/(d-1)$.
\end{enumerate}
(Note that for small dimensions $d$ not all the possibilities
enumerated above can occur. For instance for $d=2$, the only
possible values are $I_d=\pm 2$.) 
Thus for all possible choices of $r,s,t,u$, $I_d(\mbox{local realism}) 
\leq 2$. This
concludes the proof.

Let us now consider the maximum value that can be attained for the
Bell expressions $I_d$ for quantum measurements on an entangled
quantum state. We have carried out a numerical search for the optimal
measurements. It turns out that the best measurements that we have
found numerically give the same value as the measurements described in 
\cite{2}. We do not have a proof that these measurements are optimal,
but our numerical work and the numerical work that inspired \cite{2}
suggests that this is the case. 

We therefore first recall the state
state and the measurement 
described in
\cite{2}.  The quantum state is the maximally
entangled state of two d-dimensional systems
\begin{equation}
\psi = {1 \over \sqrt{d}} \sum_{j=0}^{d-1} |j\rangle_A \otimes
|j\rangle_B \ .
\label{psi}
\end{equation} 
The measurements is carried out in 3 steps. First Alice 
and Bob give each of the states $|j\rangle$ a variable phase, $e^{i
  \phi_a(j)}$ for Alice and $e^{i
  \varphi_b(j)}$ for Bob, which depends on the measurement they want to
carry out.  The state thus becomes
\begin{equation}
\psi = {1 \over \sqrt{d}} \sum_{j=0}^{d-1}  e^{i
  \phi_a(j)} e^{i \varphi_b(j)}
|j\rangle_A \otimes
|j\rangle_B \ .
\end{equation} 
where 
$\phi_1(j) = {2 \pi \over d} \alpha_1 j$, 
$\phi_2(j) = {2 \pi \over d} \alpha_2 j$,
$\varphi_1(j) = {2 \pi \over d} \beta_1 j$
and  
$\varphi_2(j) = {2 \pi \over d} \beta_2 j$ 
with 
$\alpha_1 =0$, $\alpha_2= 1/2$, $\beta_1=1/4$ and $\beta_2=
-1/4$.
The second step consists of each party 
carrying out a discrete Fourier transform
to bring the state to the form
\begin{eqnarray}
\psi &=& {1 \over {d}^{3/2}} \sum_{j,k,l=0}^{d-1}  \exp\left[i\left(
  \phi_a(j) +\varphi_b(j) +{2 \pi \over d} j (k - l)
\right)\right]\nonumber\\
& &\quad\quad
|k\rangle_A \otimes
|l\rangle_B \ .
\end{eqnarray} 
The final step is for Alice to measure the $k$ basis and Bob to
measure the $l$ basis. Thus the joint probabilities are
\begin{eqnarray}
& &P_{QM}(A_a=k,B_b=l)\nonumber\\
 &=& {1 \over d^3} | \sum_{j=0}^{d-1}
\exp \left[i {2 \pi j\over d} ( k - l + \alpha_a + \beta_b)\right]|^2
\nonumber\\
 &=& {1 \over d^3} {\sin^2[\pi (k-l+\alpha_a + \beta_b)]
\over
\sin^2[\pi (k-l+\alpha_a + \beta_b)/d]}
 \nonumber\\
 &=& {1 \over 2  d^3 \sin^2[\pi (k-l+\alpha_a + \beta_b)/d]}
\label{Pjoint}
\end{eqnarray}
where in the last line we have used the values of $\alpha_a$ and
$\beta_b$ given above.

Equation (\ref{Pjoint}) shows that 
these joint probabilities have several symmetries. First of all we
have the relation
\begin{eqnarray}
P_{QM}(A_a=k,B_b=l) = P_{QM}(A_a=k+c,B_b=l+c)
\nonumber
\end{eqnarray}
for all integers $c$. 
This symmetry property justifies us considering, as in (\ref{modulo}), 
only the probabilities 
that $A_a$ and $B_b$ differ
by a given constant integer $c$:
\begin{eqnarray}
P_{QM}(A_a=B_b+c) &=& \sum_{j=0}^{d-1} P_{QM}(A_a=j+c,B_b=j)\nonumber\\
 &=& 
d P_{QM}(A_a=c, B_b=0)\ .
\end{eqnarray}
Furthermore we have the relation
\begin{eqnarray}
&P_{QM}(A_1=B_1+c) = P_{QM}(B_1= A_2+c+1) &\nonumber\\
&= P_{QM}(A_2=B_2+c) = 
P_{QM}(B_2= A_1+c)&  \ .
\label{correlP}
\end{eqnarray} 
Using eqs. (\ref{Pjoint} to \ref{correlP}) 
we can rank these probabilities by decreasing
order.
Let us denote 
\begin{equation}
q_c = P_{QM}(A_1 = B_1 + c)
={1 / \left( 2  d^3 \sin^2[\pi (c + 1/4)/d] \right)}
\ .
\nonumber
\end{equation}
Then we have
\begin{equation}
\label{order}
q_0 > q_{-1} > q_1 > q_{-2} > q_2 > \ldots > q_{-[d/2]} 
\ \large ( > q_{[d/2]} \large )
\nonumber
\end{equation}
where $[x]$ denotes the integer part of $x$ and the last term between
parenthesis occurs only for odd dimension $d$.
This suggests that the quantum probabilities violate the
constraints imposed by local variable theories. 
Indeed the probabilities in
(\ref{correlP}) are maximized by taking $c=0$, but then the 4
relations that appear in (\ref{correlP}) are incompatible with local
realism. In fact replacing the above probabilities in the expression
(\ref{I}) yields a value $I_{QM} = 4 d q_0 > 3$ for all dimensions
$d$. 

However a stronger violation is obtained if instead of using the Bell
expression $I$, one uses the Bell expressions $I_d$. In fact for a d
dimensional quantum systems, one can use all the Bell expressions
$I_k$ for $k \leq d$, but the strongest violation is obtained by using 
the Bell expression $I_d$. This value, denoted $I_d(QM)$, is given by 
\begin{eqnarray}
I_d (QM) = 4 d
\sum_{k=0}^{[d/2] -1} \left ( 1 - { 2 k \over d-1} \right)
(q_k - q_{-(k+1)})\ .
\label{IdQM}
\end{eqnarray}
For instance we find
\begin{eqnarray}
I_3(QM) &=& 4 /\left( -9 + 6 \sqrt{3}\right) \simeq 2.87293  \ ,\nonumber\\
I_4(QM) &=& {2 \over 3} \left(\sqrt{2} + \sqrt{10 -\sqrt{2}}\right)   
\simeq 2.89624  \ ,\nonumber\\
\lim_{d\to \infty} I_d(QM) &=& {2 \over \pi^2} 
\sum_{k=0}^{\infty} {1 \over (k+1/4)^2} - {1 \over (k+3/4)^2} \nonumber\\
&=& {32\ \mbox{Catalan} / \pi^2} \simeq 2.6981
\nonumber
\end{eqnarray}
where $\mbox{Catalan}\simeq 0.9159$ is Catalan's constant.

In the presence of uncolored noise the quantum state becomes
\begin{equation}
\rho = p |\psi\rangle \langle \psi | + (1-p) {\openone \over d^2} 
\nonumber
\end{equation}
where $p$ is the probability that the state is unaffected by noise.
The value of the Bell inequality for the state $\rho$ is
\begin{equation}
I_d(\rho) = p I_d (QM)
\nonumber
\end{equation}
Hence the Bell inequality $I_d$ is certainly violated if 
\begin{equation}
p > {2 \over I_d(QM) } = p^{min}_d \ .
\label{pmin}
\end{equation}
(If there is a quantum measurement giving a value of $I_d$ greater
than that given by eq. (\ref{IdQM}), then of course the Bell
inequality would be violated with even more noise. This remark applies 
to the various $p^{min}$ below).

As a function of $d$ one finds that $p^{min}_d$ is a decreasing
function of $d$. For instance:
\begin{eqnarray}
p^{min}_3 &=& (6 \sqrt{3} - 9 ) /2 \simeq 0.69615  \nonumber\\
p^{min}_4 &=& 3 / (\sqrt{2} + \sqrt{10 -\sqrt{2}}) \simeq
0.69055 \nonumber\\
\lim_{d\to  \infty}p_d^{min}(d) &=& 
{\pi^2 / \left( 16\  \mbox{Catalan} \right)}
\simeq
0.67344\nonumber
\end{eqnarray}
For $d=3$ this reproduces the analytical result of \cite{3}. And
combining eqs. (\ref{IdQM}) and (\ref{pmin}) 
reproduces the numerical results of \cite{2} for all dimensions
($2\leq d\leq 16$) for which a numerical optimization was carried out.

In summary our reformulation of Bell inequalities in terms of a
logical constraint local variable theories must satisfy has provided
the basis for constructing a large family of Bell inequalities for
systems of large dimension. The numerical work of \cite{1,2} and a
numerical search of our own suggest that these Bell inequalities are
optimal in the same sense that the CHSH inequality is optimal for 2
dimensional systems, namely both the resistance to noise and the
amount by which the inequality are violated are maximal. For this
reason we hope that the Bell inequalities presented here will have as
much interest for physicists studying entanglement of systems of large 
dimensionality as the CHSH inequalities have had for bidimensional systems.

{\bf Note:} while completing this paper we learned of a Bell inequality
for qutrits\cite{4} that exhibits the same resistance to noise as that
obtained in \cite{1,2,3}.

{\bf Acknowledgments:} we
acknowledge funding by the European Union under project EQUIP
(IST-FET program).

\end{multicols}
\end{document}